\documentclass[final,5p,times,twocolumn,authoryear]{elsarticle}
\usepackage{placeins}
\usepackage{amssymb}
\usepackage{lipsum}
\usepackage{soul}
\usepackage{rotating} 

\journal{High Energy Astrophysics}
\usepackage{hyperref}
\usepackage{amsmath}
\hypersetup{
    colorlinks=true,
    linkcolor=blue,
    filecolor=magenta,      
    urlcolor=cyan,
    citecolor=blue,
    pdftitle={Overleaf},
    pdfpagemode=FullScreen, }

\newcommand{\src}{Swift J151857.0--572147 }

\begin{document}

\begin{frontmatter}


\title{Investigating the disappearance of a type-C QPO in black hole X-ray binary Swift J151857–572147 with \emph{AstroSat}}


\author[1]{Biki Ram}
\author[1]{Manoneeta Chakraborty}
\author[2]{Mayukh Pahari}

\address[1]{Department of Astronomy, Astrophysics and Space Engineering, Indian Institute of Technology Indore, Indore, 453552, India}

\address[2]{Department of physics, Indian Institute of Technology Hyderabad, Kandi-502284, Sangareddy, Telangana, India}

\begin{abstract}
We present a timing and broadband spectral study of the \textit{AstroSat} observation of the black hole low mass X-ray binary Swift J151857.0--572147, focusing on its evolution across two temporal segments, S1 and S2. During the first segment S1 lasting $\sim$ 50.6 ks, the source shows a clear type-C QPO at $\sim8.02$ Hz and a low-frequency bump (LFB) in the power density spectrum. By the later segment S2, the strength of the QPO feature decreases significantly and is no longer consistent with a coherent QPO, along with the simultaneous disappearance of the LFB. An energy-dependent analysis reveals that both the RMS amplitude and the time lag of the QPO in S1 increase monotonically with energy, suggesting a low inclination nature of the source. The QPO and LFB centroid frequencies show a strong positive correlation across orbital segments, with a Pearson coefficient of 0.98. Our variability analysis and hardness-intensity diagram reveal a spectral state transition of the source coincident with the turn-off of the QPO and the LFB, a conclusion supported by the evolution of broadband spectral parameters. Notably, the orbit-resolved spectral analysis reveals that the non-thermal parameters, photon index, and electron temperature of the corona exhibit significant deviation as the source traverses from S1 to S2, whereas the thermal parameters scatter without any systematic variation between the two segments. The simultaneous turn-off of the type-C QPO and the low-frequency broadband noise component, both originating from the hot inner flow, indicates a rapid reorganization of the inner accretion geometry. Our results strongly suggest a coronal origin for the QPO and demonstrate that the disappearance of both features reflects a substantial transformation of the Comptonizing region across the spectral transition.

\end{abstract}

\begin{keyword}
black hole physics \sep accretion, accretion disks \sep X-rays: binaries \sep low-mass X-ray binary stars  
\end{keyword}
\end{frontmatter}

\section{Introduction}
\label{introduction}
Low-frequency (roughly 0.1–30 Hz) quasi-periodic oscillations (LF-QPOs) are one of the most prominent timing signatures observed in black hole low-mass X-ray binaries (BH-LMXBs), offering insights into the strong gravity regime physics \textbf{\citep{2019NewAR..8501524I}}. These features have been observed in all the main spectral states of BH LMXBs outbursts
and are commonly classified into types A, B, and C based on their coherence, amplitude, and noise properties  \citep{2005ApJ...629..403C,2005A&A...440..207B, 2006ARA&A..44...49R}. Among them, type-C QPOs are the most frequently observed. They are characterized by high quality factors ($Q \geq 5-10$) and significant fractional RMS amplitudes that can exceed 15–20\%. Type-C QPOs typically show a strong correlation with spectral parameters. Their centroid frequency increases monotonically as the source evolves from the hard state to the hard-intermediate state, typically accompanied by spectral softening, an increase in the photon index, and a growing contribution from the thermal disk component \citep{2003A&A...397..729V,2012MNRAS.427..595M,2025JHEAp..4600344C}. This correlation suggests a direct link between the QPO mechanism and the geometry of the inner accretion flow. 
Despite decades of observations, the physical mechanism behind the appearance and disappearance of the feature remains one of the most debated questions in accretion physics. 

Among the proposed theoretical frameworks are the accretion–ejection instability mechanism  \citep{2022ApJ...930...18W}, instabilities of the corona–disk \citep{2004ApJ...612..988T, 2022A&A...662A.118M}, relativistic precession model \citep{1998ApJ...492L..59S, 2006ApJ...642..420S,2012MNRAS.419.2369I,10.1111}, disk instability model \textbf{\citep{2016ApJ...831...33L}}, blob-model \textbf{\citep{2009AIPC.1126..367G}}, and variable corona model \textbf{\citep{1998MNRAS.299..479L,2014MNRAS.445.2818K}}. However, none of these models can fully explain the observed characteristics of QPOs. \cite{10.1111} proposed Lense-Thirring to be the most favorable model for type-C QPOs involving a truncated disk geometry. In this framework, type-C QPOs are widely interpreted as arising from Lense–Thirring precession of the geometrically thick inner flow \citep{1998ApJ...492L..59S,1999ApJ...524L..63S,10.1111,2012MNRAS.419.2369I}. The QPO frequency is then set by the precession frequency of this hot flow, naturally increasing as the truncation radius decreases.

Understanding the geometry of the inner accretion flow around stellar-mass black holes is one of the central open problems in high-energy astrophysics, as it governs the energy output, variability, and jet activity of these systems.
Rapid turn-on and turn-off phenomena of quasi-periodic oscillations (QPOs) in black hole X-ray binaries (BHXBs) provide direct insight into rapid changes in the inner accretion geometry during state transitions. Unlike the gradual evolution of QPO centroid frequency typically observed during the hard-to-soft transition, several sources exhibit abrupt appearance or disappearance of low-frequency QPOs on timescales of seconds to minutes \citep{2003A&A...412..235N,2021ApJ...911..127S,2016ApJ...823...67S}. Such events are often accompanied by a concurrent significant evolution of the broadband noise component in the power density spectrum (PDS), suggesting a global reconfiguration of the variability-producing region.
Multiple detections of such rapid QPO transitions have previously been reported. In GX 339–4, the transitions of QPOs, including their rapid appearance and disappearance, were reported by \citet{1991ApJ...383..784M, 2003A&A...412..235N, 2011MNRAS.418.2292M}. Similar rapid transitions were observed in XTE J1550–564 \citet{2001ApJS..132..377H, 2016ApJ...823...67S}, XTE J1859+226 \citet{2004A&A...426..587C, 2013ApJ...775...28S}, and H1743–322 \citet{2005ApJ...623..383H, 2021ApJ...911..127S}. Swift J1658.2–4242 displayed a striking example where a type-C QPO appeared in a low-flux state and disappeared in a higher-flux state within $\sim$40 s \citep{2019ApJ...879...93X}, while \textit{AstroSat} observations revealed correlated spectral changes \citep{2019ApJ...887..101J}. The disappearance of the QPO has been observed for decades, but the underlying physical mechanism remains unclear.

Notably, during type-C to type-B QPO transitions, the disappearance of the type-C QPO is frequently accompanied by suppression of the strong low-frequency band-limited noise component. \cite{2025A&A...699A...9J} reported that the transition from type-C to type-B QPO was associated with the simultaneous disappearance of the low-frequency broad noise component, indicating that both features likely originate from the same precessing inner flow. Similar behavior has been observed in GX 339–4 and H1743–322 \citep{2023MNRAS.521.3570Y,2021ApJ...911..127S}, where the disappearance of broadband noise coincides with the QPO suppression.
This recurring correlation suggests a physical link between the type-C QPO and the low-frequency broad noise, which is often observed as a low-frequency ``bump'' in the PDS. The type-C QPO is ascribed to the global Lense–Thirring precession of the geometrically thick inner flow, whereas the broadband noise is believed to originate from propagating mass accretion rate fluctuations within the hot inner disk \textbf{\citep{1997MNRAS.292..679L,2001MNRAS.323L..26U,10.1111,2012MNRAS.419.2369I}}. To understand the physical origin of type-C QPOs, the geometry and dynamics of the inner accretion flow, and the mechanism underlying the rapid turn-on and turn-off events observed during spectral state transitions, it is essential to study the coupled evolution of the QPO and the low-frequency broadband component. 

\subsection{The source Swift J151857.0–572147}
Swift J151857.0–572147 is a newly discovered Galactic black hole X-ray binary that underwent its first recorded outburst in March 2024. The source was initially detected by Swift/XRT and reported as GRB 20240303A \citep{2024GCN.35853....1K}, associated with the Swift Trigger 1218452 (GCN 35849). 
The refined coordinates were determined to be RA (J2000) = 15$^{\rm h}$18$^{\rm m}$57.00$^{\rm s}$ and Dec (J2000) = $-57^{\circ}21^{\prime}47.9^{\prime\prime}$. Early radio follow-up with \textit{MeerKAT} at 1.28 GHz detected emission with a flux density of $\sim$10 mJy, consistent with compact jet emission typical of X-ray binaries in the hard state  \citep{2024ATel16518....1C}. 
\textit{Swift/XRT} spectral analysis during the early phase of the outburst revealed a hard power-law spectrum with photon index $\Gamma = 1.78 \pm 0.02$ and an absorbing neutral hydrogen column density of 
$N_{\rm H} = (5.6 \pm 0.06) \times 10^{22}\,\mathrm{cm^{-2}}$ 
\citep{2024GCN.35853....1K}, consistent with a canonical hard-state black hole spectrum. 
H\,\textsc{i} absorption studies constrained the distance to lie within $4.48$--$15.64$\, kpc \citep{2024ATel16538....1B}, although the exact value remains uncertain. 
Assuming a fiducial distance of $10$\,kpc, joint spectral modeling using \textit{IXPE} (\textit{Imaging X-ray Polarimetry Explorer}) and \textit{NuSTAR} (\textit{Nuclear Spectroscopic Telescope Array}) data yielded a black hole mass estimate of $\sim 9.2$--$10.5\,M_\odot$, a spin parameter estimate of $a_* \sim 0.6$--$0.7$, and a disk inclination in the range $\sim 35^\circ$--$47^\circ$ \citep{2024ApJ...975..257M}. These parameters place Swift~J151857.0--572147 among the moderately inclined stellar-mass black hole systems.  \cite{2024ApJ...973L...7P} also performed a spectral analysis using \textit{NICER} (\textit{Neutron Star Interior Composition Explorer}), \textit{NuSTAR}, and \textit{HXMT} (\textit{Hard X-ray Modulation Telescope}) data to understand the evolution of the source during its outburst.
Moreover, using \textit{IXPE}, \cite{2024ApJ...975..257M} reported the first X-ray polarization detection from this source, indicating a non-negligible polarization degree (1.34 $\pm$ 0.27 \%),
constraining the coronal geometry and emission mechanism. This is consistent with a small corona (6$\pm$1 - 9$\pm$2 $R_S$) and a low mass outflow rate.
The low polarisation degree (PD) likely results from repeated scattering within the dense corona.
A type-C QPO has been detected by \cite{2025ApJ...987...44C} in Swift J151857.0–572147 with \textit{HXMT} data, which allows probing the inner region of the accretion disk. 
However, detailed energy-resolved timing studies, including RMS and time-lag spectra, which are essential for constraining the physical origin and evolution of the QPO, have not been conducted for this source. 
In particular, QPO analysis serves as a powerful diagnostic of the geometry, dynamics, and radiative processes in the inner region of the accretion disk, helping us constrain the evolution of the disk and the corona during state transitions. 
In this context, \textit{AstroSat}, with its broad energy coverage, large effective area, and sensitivity to rapid flux variations, is uniquely suited to characterize QPO behavior with the precision and temporal resolution required to 
trace the evolution of the inner accretion geometry, especially during a rapid state transition of the source.

In this work, we utilize simultaneous \textit{AstroSat} SXT (Soft X-ray Telescope) and LAXPC (Large Area X-ray Proportional Counter) observations of Swift J151857.0–572147 for the first time to perform a detailed timing and broadband spectral analysis during the disappearance of a type C QPO, associated with a harder to a softer state transition, and investigate its correlation with changes in spectral parameters.
This study provides observational constraints on the physical mechanism responsible for the suppression of coherent QPO modulation and the restructuring of the inner accretion flow in this newly discovered black hole transient.
\begin{figure}[ht!]

\includegraphics[scale=0.45]{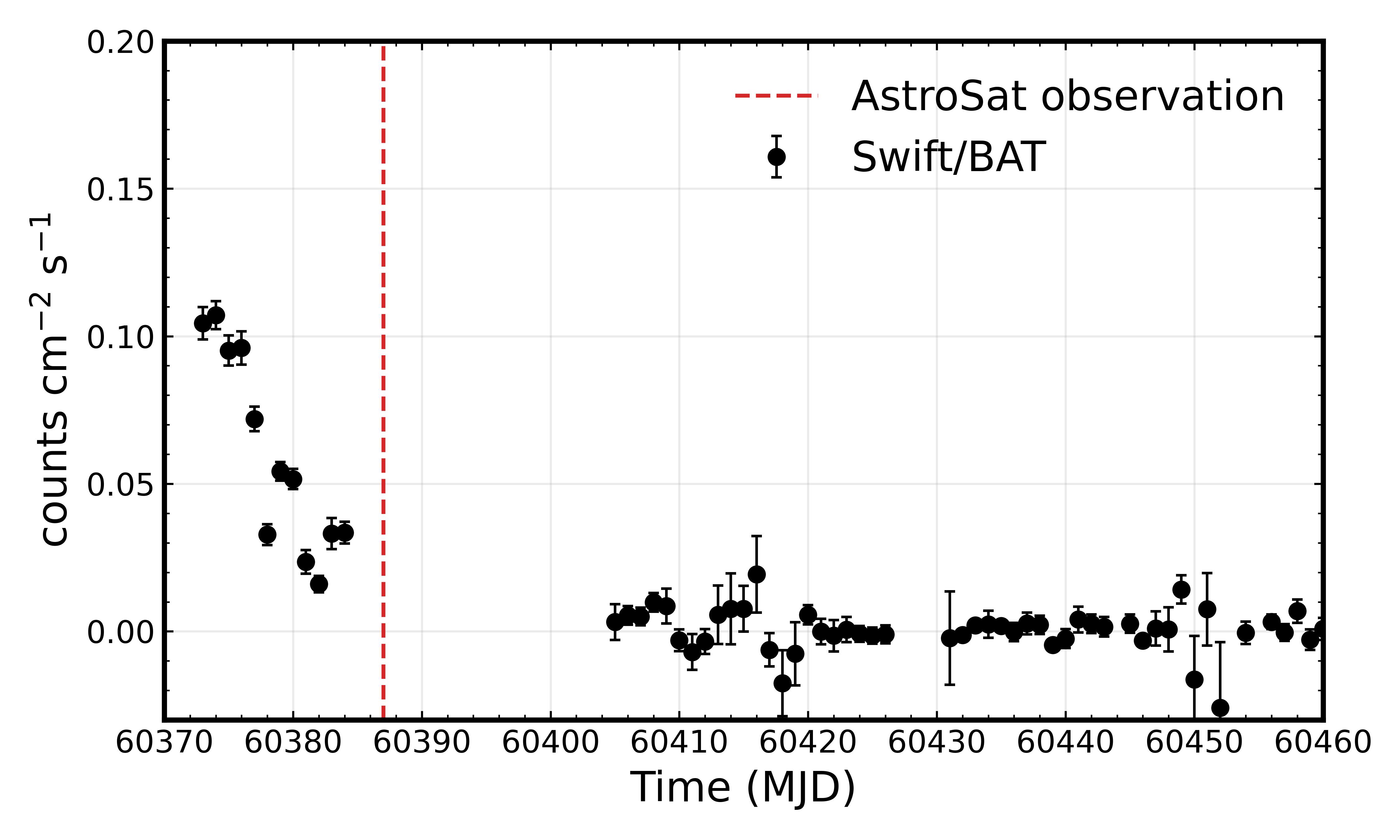}
\caption{The light curve of Swift J151857.0–572147 from Swift/BAT in
15–50 keV, with the dashed red line representing the epoch of the \textit{AstroSat} observation. \label{Fig:1}}
\end{figure}
\section{Observation and data reduction \label{Sec:2}}
In this work, we present a detailed broadband spectral–timing analysis of Swift J151857.0–572147 using \textit{AstroSat} observation (ObsID: 9000006126) conducted on 18 March 2024, starting at 17:03:27 UTC, with a total exposure of 100.99 ks.
\textit{AstroSat} is India’s first dedicated multi-wavelength space observatory for astronomical studies \citep{2017JApA...38...30A}. It carries five scientific payloads: the Large Area X-ray Proportional Counter (LAXPC), the Soft X-ray Telescope (SXT), the Ultra-Violet Imaging Telescope (UVIT), the Cadmium-Zinc-Telluride Imager (CZTI), and the Scanning Sky Monitor (SSM). In this analysis, we exclusively utilize data obtained simultaneously with the LAXPC and the SXT instruments.
\begin{figure*}[ht!]
\begin{center}
\includegraphics[scale=0.52]{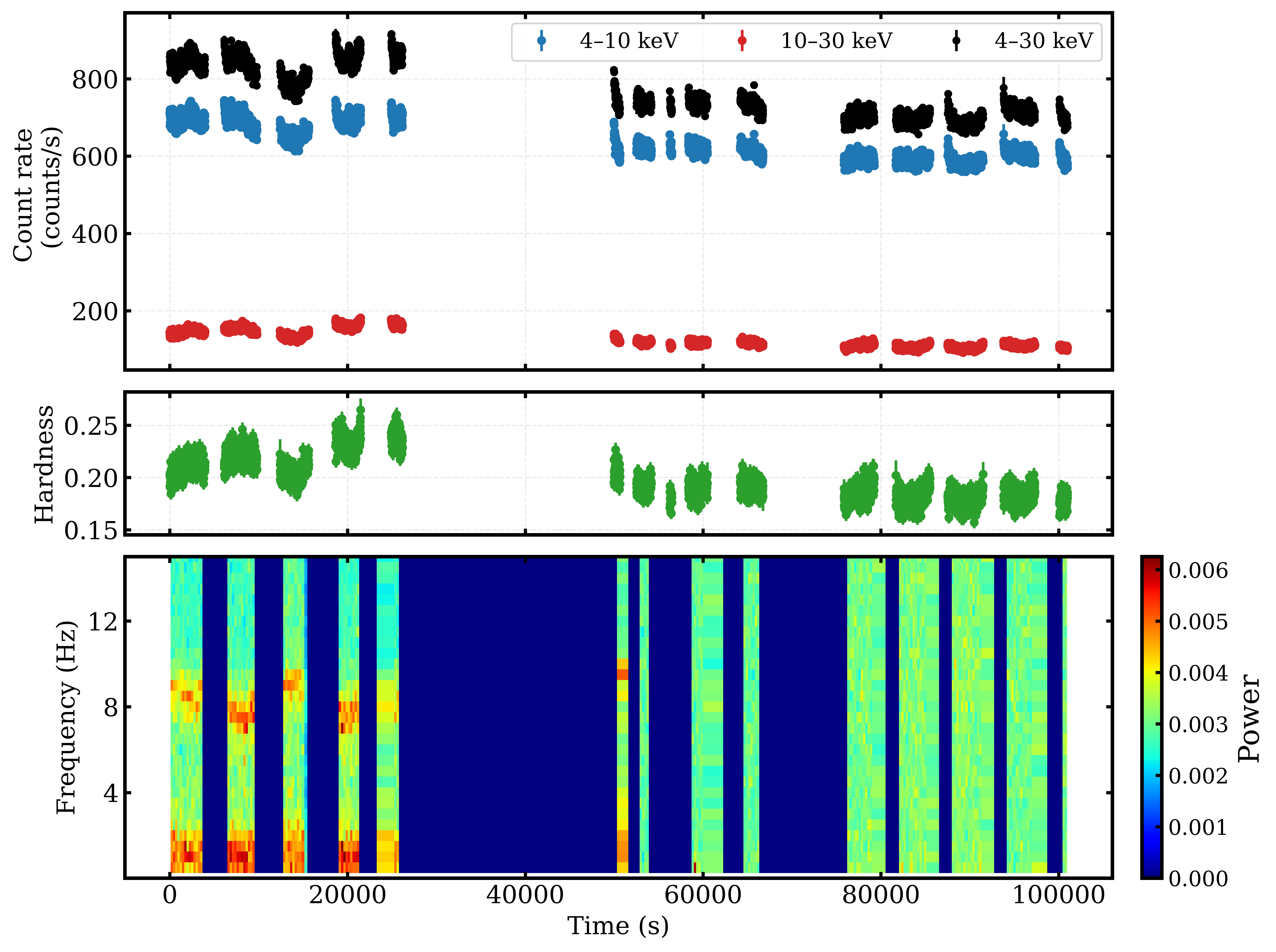}
\caption{LAXPC timing properties of the source Swift J151857.0–572147 during the \textit{AstroSat} observation. The top panel shows the background-subtracted light curves in the 4–10 keV (blue) and 10–30 (red) keV energy bands with a time bin of 10 s, along with the total count rate in 4-30 keV (black). A gradual decline in the count rate is observed throughout the observation. The middle panel displays the corresponding hardness ratio, computed using the two energy bands. The bottom panel shows the dynamic power spectrum for the entire observation, generated with the {\tt laxpc\_find\_freqlag} tool and an integration time of 256 s, spanning 0.5–16 Hz.  \label{Fig:2}}
\end{center}
\end{figure*}

\subsection{LAXPC} 
Given its large effective area and excellent timing capability, LAXPC is particularly well-suited for in-depth temporal investigations of BH-LMXBs. The instrument consists of three identical gas-filled proportional counter units—LAXPC10, LAXPC20, and LAXPC30—each filled with a gas mixture of 90\% Xenon and 10\% Methane \citep{2017CSci..113..591Y,2017JApA...38...30A}. Soon after launch, LAXPC30 suffered a gas leakage and was therefore excluded from scientific use. In addition, LAXPC10 began showing indications of gas leakage and gain instability after 2018. To ensure uniformity across all observations, we restrict our analysis to LAXPC20 data only. LAXPC provides a timing resolution of 10 $\mu$s with a detector dead time of 42 $\mu$s. It operates over an energy range of 3–80 keV and has an effective area of approximately 6000 cm$^2$/keV at 15 keV \citep{2017CSci..113..591Y,2017JApA...38...30A}. The energy resolution of the instrument is about 15–20\% at 30 keV. In this work, we restrict the analysis to energies below 30 keV, as background contributions become increasingly dominant at higher energies, thereby compromising the reliability of the source signal. All analyses are performed using event mode data from LAXPC20.

The raw level-1 data were obtained (Obs ID: 9000006126) from the \textit{AstroSat} Data Archive\footnote[2]{https://astrobrowse.issdc.gov.in/astroarchive/archive/Home.jsp}
 and processed into level-2 products using the LAXPC FORMAT-A software\footnote[1]{http://astrosat-ssc.iucaa.in/laxpcData}. For each observation, orbit-wise data files were merged using the {\tt laxpc\_make\_filelist} tool. Good Time Interval (GTI) files were then generated with {\tt laxpc\_make\_stdgti} to remove time intervals affected by Earth occultation and passages through the South Atlantic Anomaly (SAA). Using the filtered GTIs, light curves were produced with the {\tt laxpc\_make\_lightcurve} task. In several cases, residual artifacts remained in the light curves even after standard GTI filtering. These features were removed by manually refining the GTI files to exclude the affected time intervals. The revised GTIs were subsequently applied to generate source spectra and background spectra using the {\tt laxpc\_make\_spectra} and {\tt laxpc\_make\_backspectra} utilities, respectively. For timing analysis, power density spectra were computed using the {\tt laxpc\_find\_freqlag} tool. We performed the timing analysis using LAXPC only because of its high timing resolution. 

\subsection{SXT}
The Soft X-ray Telescope (SXT) onboard \emph{AstroSat} operates in the 0.3–8.0 keV energy range and has an effective area of approximately 90 cm$^{2}$ at 1.5 keV \citep{2017JApA...38...29S}. In this work, we utilized level-2 SXT data obtained in the Photon Counting (PC) mode, which provides a temporal resolution of 2.3775 s. The cleaned event files from individual orbits corresponding to each observation were combined using the {\tt SXTEVTMERGERTOOL}.
To mitigate the pile-up effect, we excluded the central region of the CCD and extracted source events from an annular region with inner and outer radii of $2'$ and $12'$, respectively. 
The images, light curves, and spectra were extracted using {\tt XSELECT} (v2.4m). To account for the modified extraction regions, we generated appropriate ancillary response files using the updated {\tt SXTARFModule} (v03), employing the default SXT ancillary response file (sxt\_pc\_excl00\_v04a\_20240917.arf). The response matrix file sxt\_pc\_mat\_g0to12.rmf was used for this observation. Additionally, the background sky spectrum file SkyBkg\_comb\_EL3p5\_Cl\_Rd16p0\_v01.pha, provided by the SXT team, was incorporated during the spectral analysis.

\section{Results}
\subsection{\bf Light curve, dynamic power spectra, and HID}
The light curve of Swift J151857.0–572147 during the 2024 outburst using \textit{Swift/BAT} is presented in Figure~\ref{Fig:1} with black data points. The dashed red line represents the epoch of the \textit{AstroSat} observation.
The light curve using different energy bands of LAXPC is plotted in the upper panel of Figure~\ref{Fig:2}, considering a bin size of 10 seconds.
We have plotted the hardness ratio considering 4-10 keV as the soft band and 10-30 keV as the hard band in the middle panel of Figure~\ref{Fig:2}. The third panel of the figure represents the dynamic power spectra for the whole observation, using {\tt laxpc\_dynpower} tool, considering 256 s of individual segments of the light curve with a minimum frequency of 0.5 Hz and a maximum frequency of 16 Hz. In the dynamic power density spectra, we have observed a QPO near 8 Hz and also a low-frequency bump near 1 Hz. But these features were only present in the first 6 orbits of the \textit{AstroSat} observation (third panel of Figure~\ref{Fig:2}). Thus, to understand the behavior of the QPO, we have divided the light curve into two segments, S1 and S2. The S1 segment represents the first 6 orbits of the light curve, and S2 represents the last 9 orbits. 

\begin{figure}[ht!]
\includegraphics[scale=0.4]{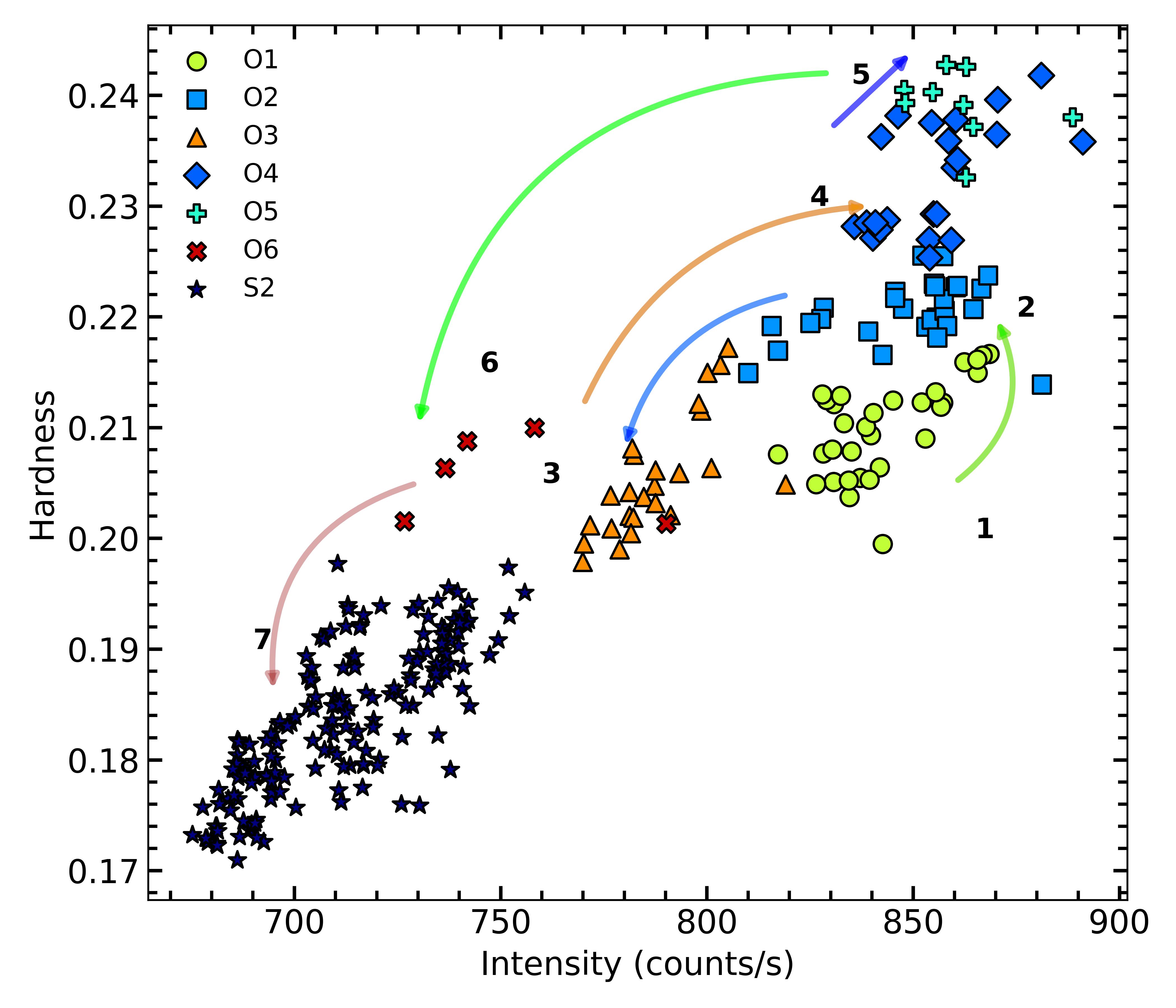}
\caption{Hardness intensity diagram of Swift J151857.0–572147 during the 2024 outburst. The hardness is defined as the count rate in the hard band (10–30 keV) normalized by the count rate in the
soft band (4–10 keV), and the intensity is the total count rate
in the 4–30 keV energy range. The colored data points represent the S1 segment, and the black data points represent the S2 segment. Different markers indicate different orbits of S1. The arrows depict the chronology of the source evolution in the HID.
\label{Fig:3}}
\end{figure}

The hardness intensity diagram (HID) for the LAXPC observation of the 2024 outburst of Swift J151857.0–572147 is shown in Figure~\ref{Fig:3}. The hardness is defined as the count rate in the hard band (10–30 keV) normalized by the count rate in the soft band (4–10 keV), and the intensity is the total count rate in the 4–30 keV energy range. The S1 segment is represented by colored data points, while S2 is represented by black star symbols. In S1, the various marker styles, in different colors, identify the individual orbits (O1–O6).  
The arrows show the chronology of the source evolution in the HID. The HID shows a clear spectral evolution during the S1 segment and from S1 to S2. The source has a relatively high intensity ($\sim$840–870 counts/s) and moderate hardness ($\sim$0.20–0.22) in orbit O1, followed by a slight increase in hardness in O2. From O2 to O3, the source gradually moves towards lower intensity and lower hardness. At O4 and O5, the source reaches the hardest edge of the S1 track (hardness $\sim$ 0.23-0.24) with maximum intensity ($\sim$ 850-900 counts/s). In orbit O6, the source shows a substantial decrease in intensity ($\sim$ 740–760 counts/s) and moderate hardness ($\sim$ 0.20–0.21), suggesting a move away from the earlier track. Beyond this, in the S2 segment, the QPO disappears, as seen in the dynamic power spectra (lowest panel of Figure~\ref{Fig:2}).  The S2 segment (black stars) occupies a distinctly different region of the HID, with lower intensities ($\sim$ 680–760 counts s$^{-1}$) and softer spectra (hardness $\sim$ 0.17-0.20). 

\begin{figure}[ht!]
\begin{center}
\includegraphics[scale=0.45]{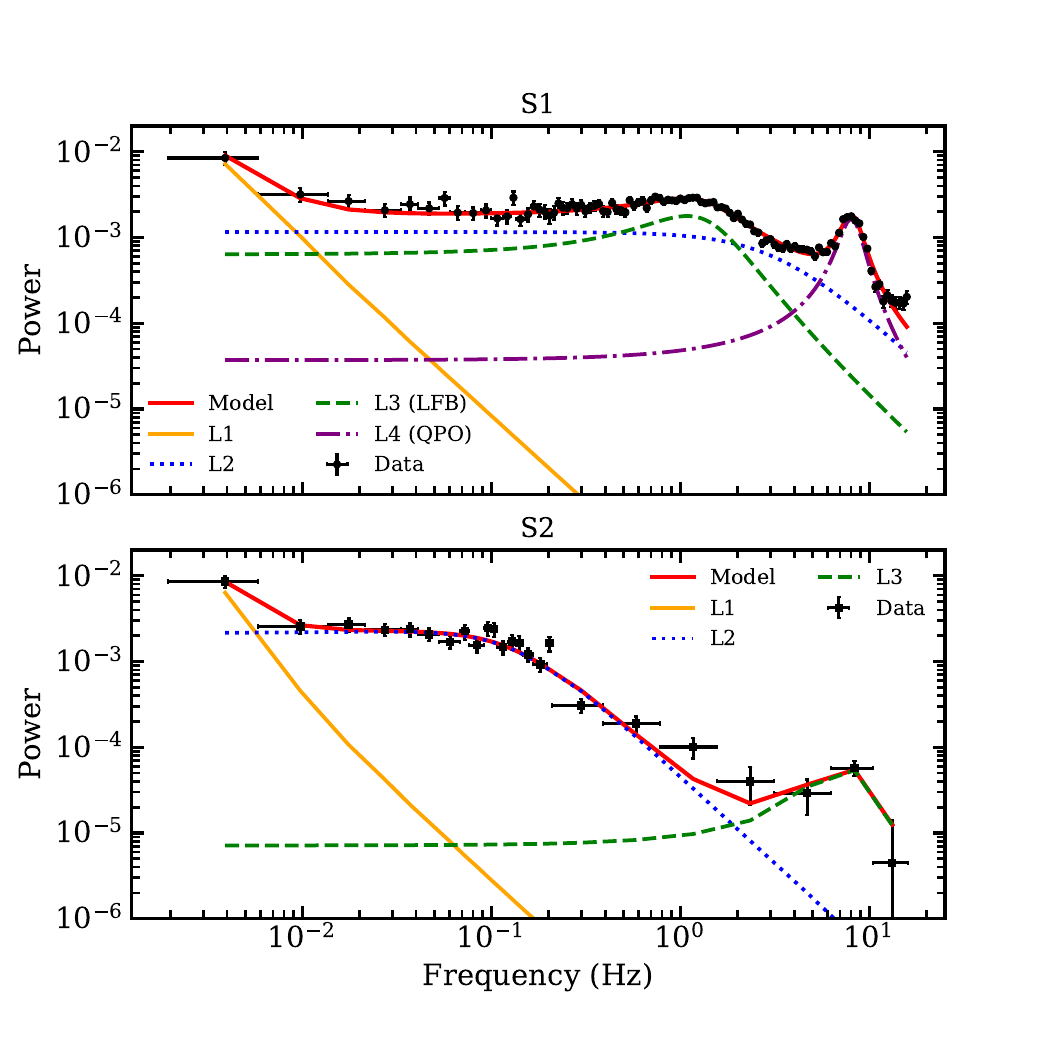}
\caption{ RMS normalized power spectra for S1 (upper panel) and S2 (lower panel) segments, fitted with multiple lorentzians as shown with the different colors. Adopting an F-test threshold criterion of 3$\sigma$, the power spectra of S1 and S2 were fitted with 4 and 3 lorentzians, respectively.  \label{Fig:4}}
\end{center}
\end{figure}

We have generated the power spectra for both the S1 and S2 segments separately using the {\tt laxpc\_find\_freqlag} tool, considering a Nyquist frequency of 16 Hz and a frequency resolution of 0.0039 Hz for the energy range of 4.0-30.0 keV. We fitted the power spectra with a combination of multiple lorentzians. A Lorentzian component was included in the final model only if its addition significantly improved the fit, verified using the F-test at a 3$\sigma$ confidence level. We also checked the power spectra in the high-frequency range and found no significant feature above 16 Hz, similar to what was reported in \citet{2025ApJ...987...44C}. 
The top panel of Figure~\ref{Fig:4} represents the fitted power spectrum of the S1 segment, where the clear presence of a type C QPO and a low-frequency bump (LFB) near 1 Hz is observed. The modeled power spectrum for the S2 segment is shown in the bottom panel of Figure~\ref{Fig:4}. We calculated the hue values for both S1 and S2 segments using the method described by \cite{2015MNRAS.448.3339H} and found a significant change in the hue value from 177$^\circ$ to 13$^\circ$, which suggests the source has undergone a significant state transition from a harder to a softer state. 

\vspace{0.1cm}

\subsection{\bf Energy-dependent timing analysis}
\begin{figure}[ht!]
\includegraphics[scale=0.43]{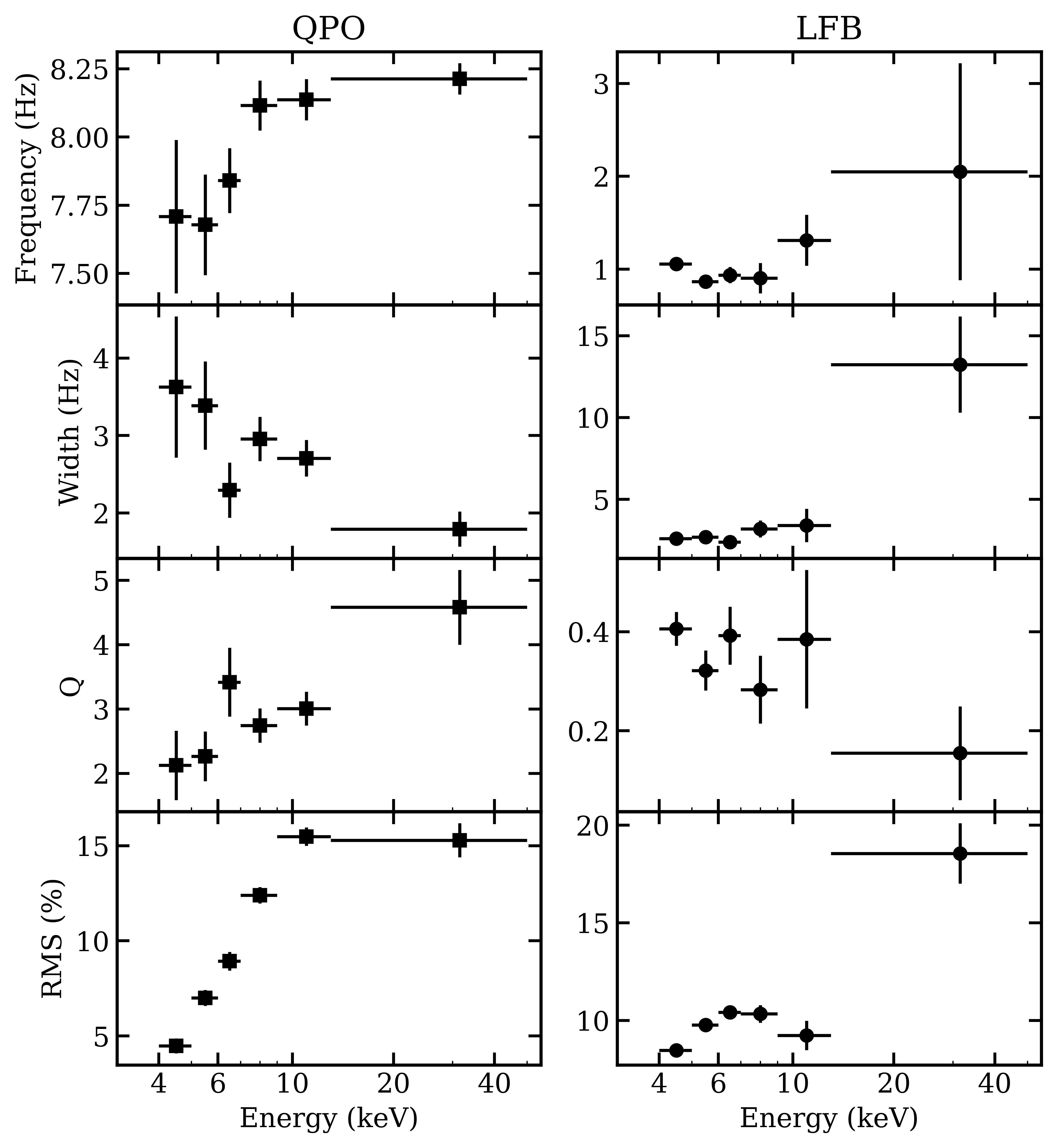}
\caption{Energy dependence of the frequency, width, $Q$ factor, and RMS amplitude of the QPO (left panel) and the LFB (right panel). \label{Fig:5}}
\end{figure}

To investigate the nature of the QPO and the LFB, we performed an energy-dependent variability analysis. The analysis was carried out using the following energy bands: 4.0–5.0 keV, 5.0–6.0 keV, 6.0–7.0 keV, 7.0–9.0 keV, 9.0–13.0 keV, 13.0–50.0 keV. For each energy band, power density spectra were generated for the S1 segment and modeled using a combination of Lorentzian components.
The RMS spectrum was derived from the area under the curve of the Lorentzian corresponding to the QPO and the LFB. The evolution of the parameters of these two features can be seen in Figure~\ref{Fig:5}. The errors on the parameters correspond to 1$\sigma$  confidence intervals. The QPO frequency and $Q$ factor showed an increase with energy. The RMS amplitude also exhibits an increasing trend with energy, from $\sim 5$\% in the 4.0–5.0 keV band to $\sim 15$\% at 10 keV. This increasing behavior persists up to 10 keV, beyond which the RMS amplitude shows an indication of a saturation at 15\%.
We also examined the energy dependence of the low-frequency bump using a method similar to that discussed earlier. The RMS spectra of the low-frequency bump show a non-monotonic behavior (lower right panel of Figure~\ref{Fig:5}). Initially, it increases across the first three energy ranges, then decreases slightly before increasing again in the last energy range. 
\begin{figure}
\begin{center}
\includegraphics[scale=0.35]{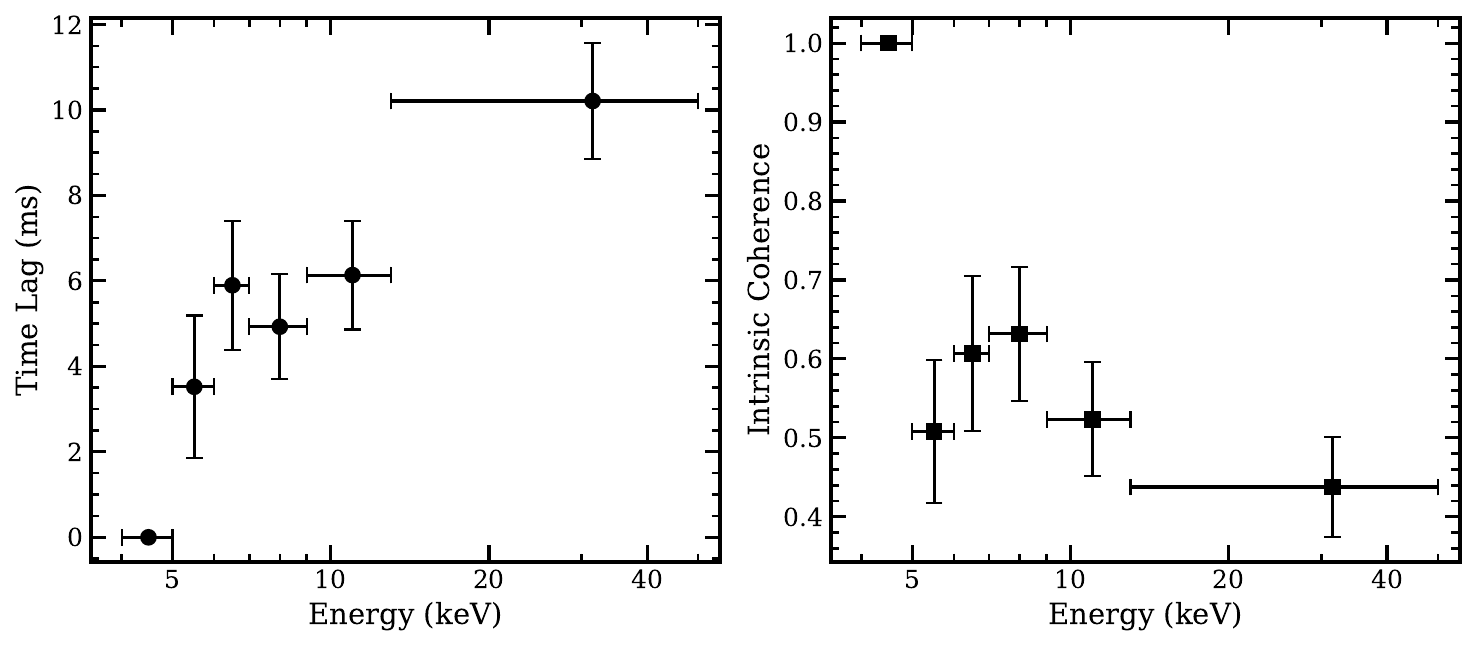}
\caption{Time-lag and coherence spectra of the $\sim 8.0$ Hz QPO obtained with the 4.0-5.0 keV band used as the reference band. \label{Fig:6}}
\end{center}
\end{figure}

To compute the time lag and coherence at the QPO frequency, we used the subroutine {\tt laxpc\_find\_freqlag}, which requires the centroid frequency feature ($\nu$) and the frequency resolution ($\Delta$ $\nu$) as inputs, along with the selected energy bands and the corresponding GTIs. The frequency resolution $\Delta$ $\nu$ was determined by the width of the QPO lorentzian. 
For coherence and time-lag estimation, the 4.0–5.0 keV energy band was chosen as the reference band. 
The time-lag spectrum also shows a monotonic increase with energy (Figure~\ref{Fig:6}). The measured time lags rise from about $\sim$4 ms at the lowest (5.0-6.0 keV) energy band to $\sim$ 10 ms in the 13–50 keV energy range.

\subsection{\bf Orbit-wise timing analysis \label{sec:3.3}}
\begin{figure}
\vspace*{-1cm}
\begin{center}
\includegraphics[scale=0.55]{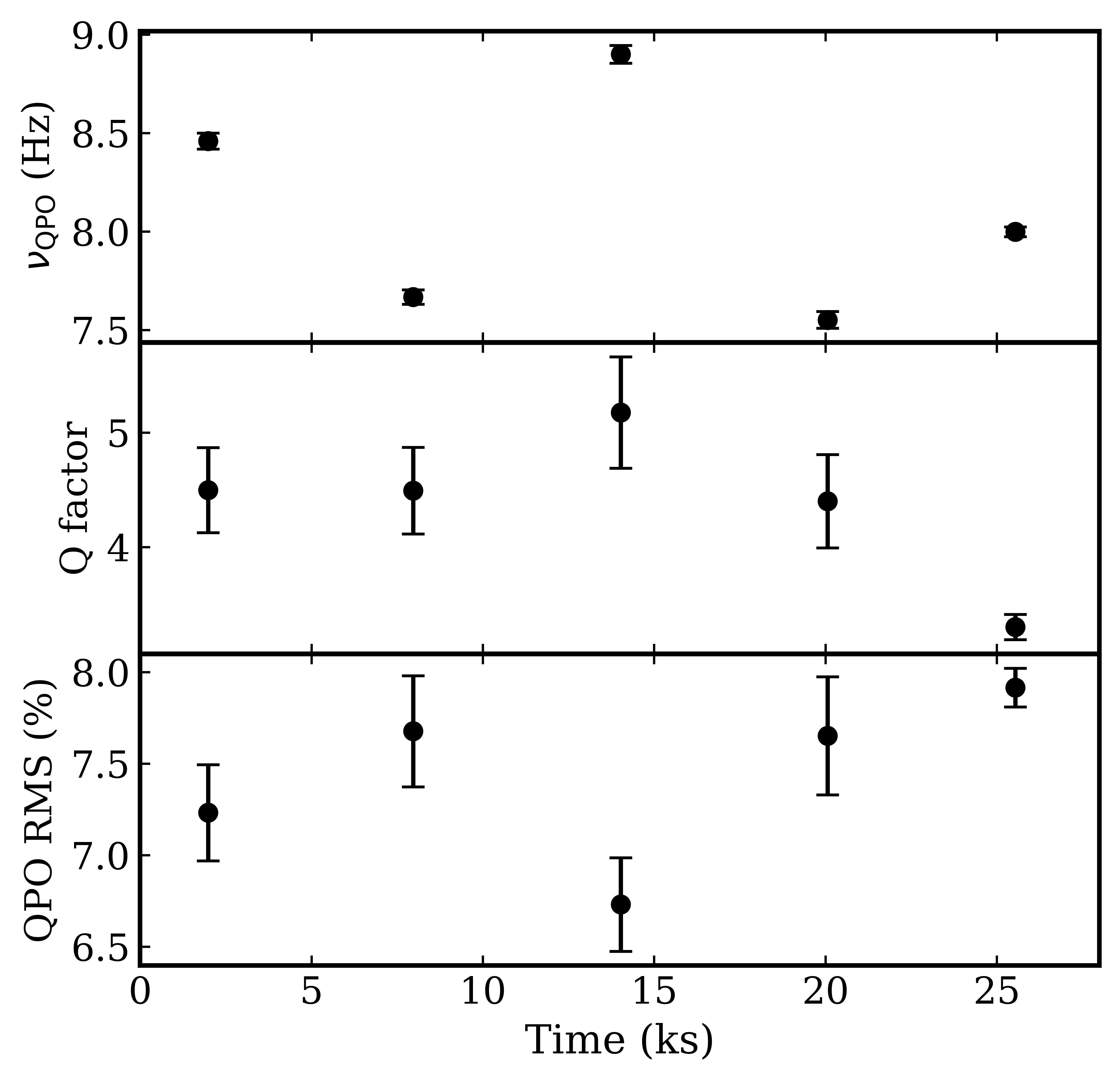}
\caption{ Orbit-resolved evolution of the QPO centroid frequency ($\nu$), quality factor (Q), and fractional RMS amplitude during the S1 observation. Error bars represent the 1$\sigma$ uncertainties obtained from the Lorentzian fits to the power density spectra.  \label{Fig:7}}
\end{center}
\end{figure}

To investigate the evolution of the QPO along with the evolution of the low-frequency bump, we further divided the S1 segment into its individual orbits and performed the timing and joint spectral analysis for each orbit. Initially, the S1 segment was divided into 6 individual orbit segments. 
However, simultaneous SXT data were unavailable for the sixth orbit; therefore, the orbit was excluded from further analysis. We continued with the first five orbits and identified them as O1, O2, O3, O4, and O5. 
We generated the power spectra for all five segments, considering a Nyquist frequency of 16 Hz and a frequency resolution of 0.0039 Hz, and fitted them with multiple Lorentzians. We observe a clear evolution in the QPO centroid frequency, Q factor, and RMS across the five orbital segments  ( Figure~\ref{Fig:7}).
\begin{figure}
\begin{center}
\includegraphics[scale=0.62]{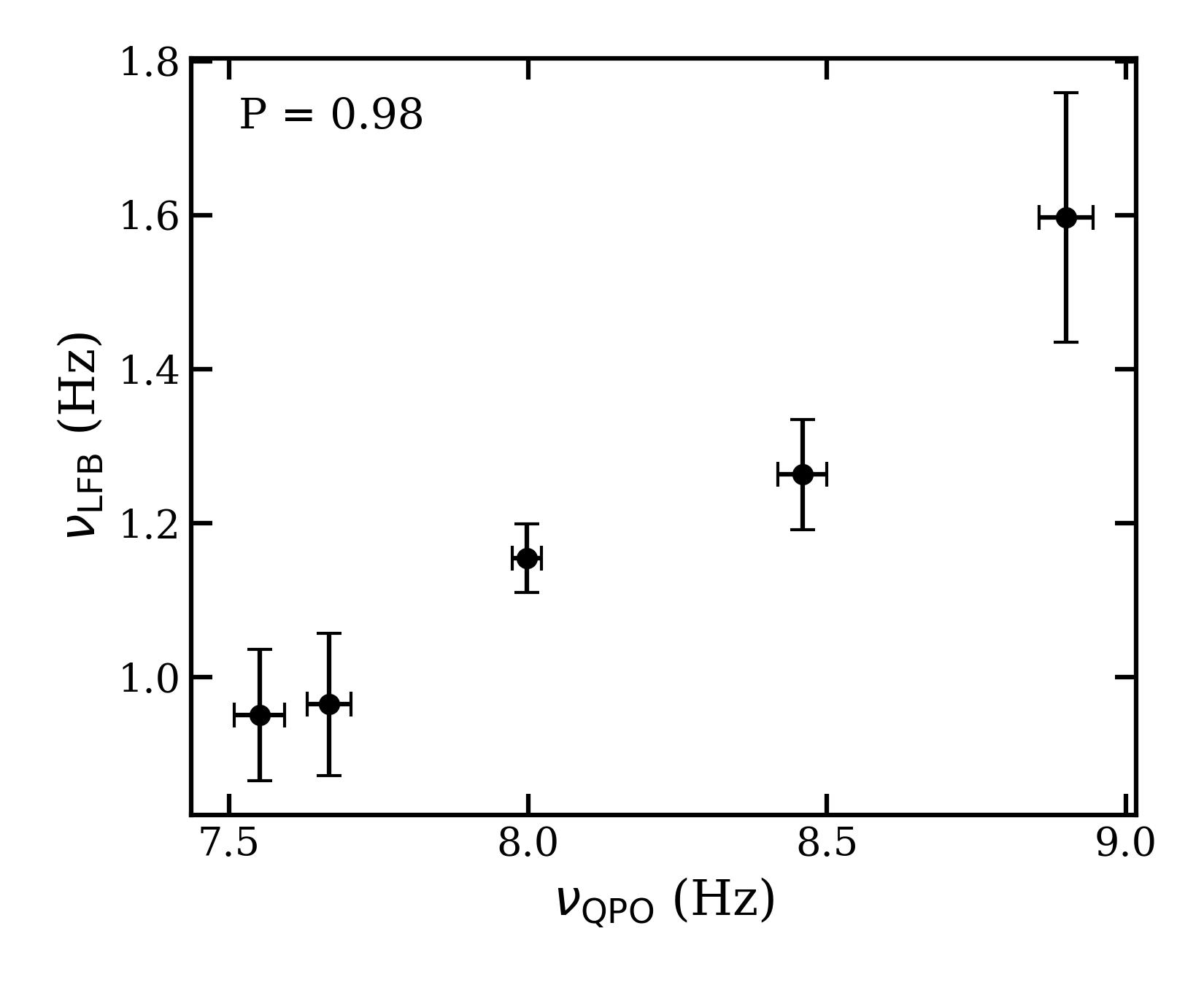}
\caption{Correlated evolution of the centroid frequency of the QPO and the low-frequency bump across five orbital segments of S1. The Pearson correlation coefficient of 0.98 is shown in the plot legend. \label{Fig:8}}
\end{center}
\end{figure}
The QPO RMS evolves between 6.71\%  and 7.94\% (O5) as shown in the third panel of Figure~\ref{Fig:7}, indicating a strengthening of the variability towards the latter segments.
Similarly, the low-frequency bump (LFB) also shows significant variability in its timing properties. The corresponding LFB RMS evolves from 5.00\% (O1) to 8.12\% (O2), drops to 3.74\% (O3), increases to 8.37\% (O4), and decreases to 5.20\% (O5). The low-frequency bump frequency also varied between orbits along with the QPO. The correlated evolution of the frequency of both these features can be seen in Figure~\ref{Fig:8}. A strong positive correlation is observed with a Pearson correlation coefficient of 0.98.

\begin{table}[h!]
\centering
\begin{tabular}{cccc}
\hline
\hline
\textbf{Parameter} & &S1 & S2 \\
\hline
Diskbb& $T_{\rm in}$ (keV) & 1.08$\pm$0.01 & 1.10$\pm$ 0.01\\\\
&$R_{\rm in}$ (R$_{g}$)  &  {\bf $2.90_{-0.02}^{+0.02}$} &  {\bf $2.95_{-0.02}^{+0.02}$} \\\\\\
nthComp& $\Gamma$   & $1.86_{-0.05}^{+0.05}$ &  $1.45_{-0.07}^{+0.04}$ \\\\

&$kT_{\rm e}$ (keV) & $6.4_{-0.3}^{+0.4}$ &  $4.5_{-0.1}^{+0.1}$ \\\\
&norm &$0.43_{-0.04}^{+0.04}$ &$0.11_{-0.02}^{+0.02}$\\\\\\
Flux$^{x}$ & Thermal&  
 $41.7_{-0.5}^{+0.5}$ 
& $46.9_{-0.2}^{+0.2}$\\\\
& Non-thermal  
& $47.9_{-0.4}^{+0.4}$ 
& $28.1_{-0.1}^{+0.1}$\\\\
&Total   
& $88.7_{-0.5}^{+0.5}$ 
& $76.1_{-0.5}^{+0.5}$ \\\\\hline

Reduced $\chi^2$ (dof)& &1.08 (647)&1.38 (655)\\
\hline
\end{tabular}
\caption{Best fitted spectral parameters corresponding to the best-fitted model. \label{Tab:1}}
\begin{flushleft}
    \footnotesize{$T_{in}$: inner disk temperature}\\
     \footnotesize{ $R_{\rm in}$: Inner disk radius calculated from disk blackbody normalisation}\\
    \footnotesize{$\Gamma$: power law index}\\
    \footnotesize{$kT_e$: electron temperature}\\
    \footnotesize{$^x$  Flux level in units of $10^{-10}$ erg s$^{-1}$  cm$^{-2}$ in the energy range of 4 - 25 keV}
\end{flushleft}
\label{tab:spectral_comparison}
\end{table}

\subsection{\bf Spectral analysis \label{sec:3.4}} 
\begin{figure}
\vspace*{-1cm}
\begin{center}
\includegraphics[scale=0.55]{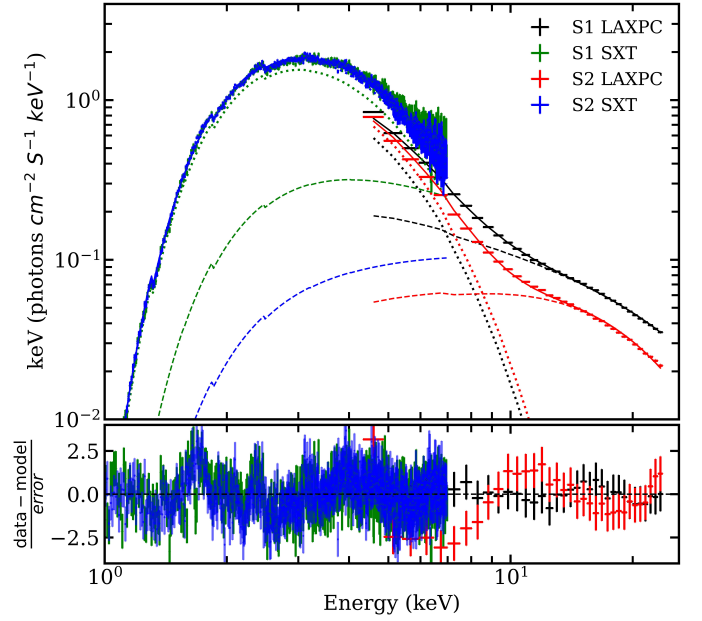}
\caption{ Fitted energy spectra for S1 segment  and for S2 segment. Different colors represent data and model components (dashed lines) from different instruments for the two segments S1 and S2.} \label{Fig:9}
\end{center}
\end{figure}

A joint spectral analysis was performed by combining the LAXPC spectrum in the 4–25 keV range and the SXT spectrum covering 0.7–7.0 keV. The analysis was carried out using { \tt XSPEC} version 12.12.0, part of the {\tt HEASOFT} package (version 6.29). Photons with energies above 25 keV were excluded due to significant background contamination in this range. A systematic uncertainty of 3\% was applied throughout the analysis as suggested by the instrument team. The photoelectric cross section and abundances were adopted from \cite{2000ApJ...542..914W}. Interstellar absorption was accounted for using the \texttt{tbabs} model.
We allowed $N_{\mathrm{H}}$ to be a free parameter during spectral fitting and obtained significant constraints on the spectral parameters for both S1 and S2 segments. In this case, $N_{\mathrm{H}}$ was estimated as $4.83 \times 10^{22} \, \mathrm{cm^{-2}}$ and $4.94 \times 10^{22} \, \mathrm{cm^{-2}}$ for S1 and S2, respectively. We subsequently chose the average of the $N_{\mathrm{H}}$ values of S1 and S2 segments, which is $4.89 \times 10^{22} \, \mathrm{cm^{-2}}$. Notably, this value lies well within the range reported in earlier studies \citep{2024ApJ...975..257M}. To fit the joint spectra, we used {\tt diskbb} to model the thermal emission and {\tt nthComp} to model the non-thermal emission.  
Since we later performed spectral analysis of individual orbital segments as well, where the signal-to-noise ratio is relatively much lower, we adopted a relatively easy-to-constrain yet physically motivated model and did not use more complex models due to sensitivity limitations.
It should be noted that our motivation here was to trace the spectro-temporal evolution over short time scales.
The best-fitting model (\texttt{TBabs × constant (diskbb + nthComp)}) parameters for the segments S1 and S2 are listed in Table~\ref{Tab:1}, along with their corresponding reduced chi-square values. Here, the uncertainties quoted are at the $1\sigma$ level. The unabsorbed thermal, non-thermal, and total flux values in the 4-25 keV range are also mentioned in the table.
Figure~\ref{Fig:9} represents the best fit of the spectrum generated by considering the simultaneous LAXPC and SXT spectrum of both S1 and S2 segments. 
The residuals seen near 1.8 keV and 2.2–2.4 keV are instrumental lines that have been reported in earlier works \citep{2023MNRAS.520.5828B}. 
We have also explored the updated, convolution-based Comptonization model, {\tt thcomp}. Both models provided consistent parameter trends and values. The only difference is in the disk normalization, where the value is slightly higher, but the difference between S1 and S2 for both cases is minute. 
We have estimated the inner disk radius ($R_{\rm in}$) from the \texttt{diskbb} normalization, following standard convention, adopting a source distance of 10 kpc and an inclination of 60$^{\circ}$ \citep{2024ApJ...975..257M}. For the black hole mass, we adopted the mean of the reported range (9.2--10.1 M$_{\odot}$) in \citet{2024ApJ...975..257M}, yielding a mass $M = 9.65$ M$_{\odot}$, which was used to compute $R_{\rm in}$ in units of the gravitational radius ($R_{g}$). 
During the transition from S1 to S2, coinciding with the disappearance of the QPO, the spectral parameters show significant evolution. 
The Comptonized component shows a pronounced change, with the photon index decreasing from $1.86 \pm 0.05$ to $ 1.45 \pm 0.07$ (Table ~\ref{Tab:1}). This is accompanied by a substantial drop in the Comptonization normalization and a decrease in electron temperature from $6.42 \pm 0.42$ keV to $4.46 \pm 0.15$ keV. 
Consistently, the flux evolution shows an increase ($\sim$12\%) in the thermal flux and a significant decrease ($\sim$41\%) in the non-thermal flux, leading to an overall reduction in the total flux.

\subsection{\bf Orbit-wise spectral analysis \label{sec:3.5}}
\begin{figure}[ht!]
\includegraphics[scale=0.37]{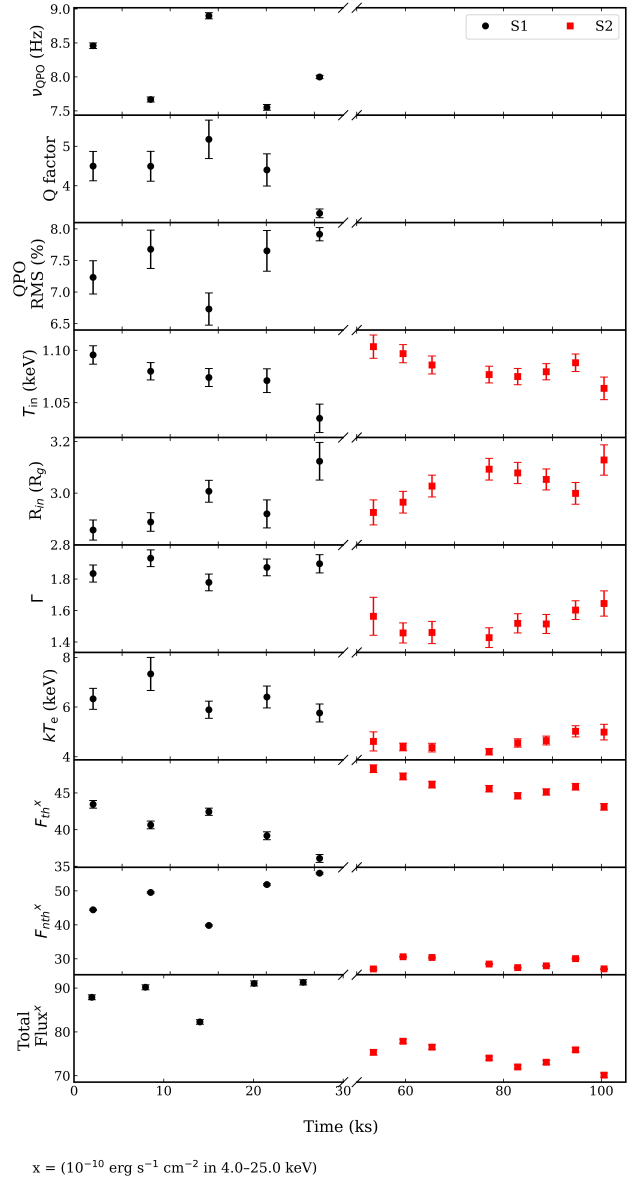}
\caption{Evolution of spectral parameters with time for all the considered orbit data in S1 (black circle) and S2 (red square) segments of the \emph{ AstroSat} observation of Swift J151857.0-572147. First panel: QPO frequency, second panel: Q factor of the QPOs, third panel: 
fractional RMS of the QPOs. fourth panel:
inner disk temperature, fifth panel: inner disk radius, sixth panel: power law index, seventh panel:
electron temperature, eighth panel: 
 thermal flux, ninth panel: non-thermal flux, tenth
panel: total flux. Note that temporal parameters related to the QPO are shown only for the S1 segment, where the QPO was present. Please note that the first three panels are the same as Figure~\ref{Fig:8} and redisplayed here for comparison with the spectral parameters.} \label{Fig:10}  
\end{figure}

We performed a joint broadband spectral analysis of the SXT and LAXPC data using the same model as described in the previous section to investigate the evolution of the accretion properties across the five segments (all belonging to S1) as mentioned in section ~\ref{sec:3.3}. The systematic error,  photoelectric cross section, and hydrogen column density are taken to be the same as discussed earlier (section~\ref{sec:3.4}). 
The inner disk temperature ($T_{\rm in}$) shows a systematic decline from $1.08\pm 0.01$ keV in O1 to $1.02 \pm 0.01$ keV in O5. Simultaneously, the \texttt{diskbb} normalization increases overall from 896 $\pm$ 21 to 1074 $\pm$ 43, with moderate intermediate fluctuations.
The photon index ($\Gamma$) and electron temperature of the Comptonizing plasma ($kT_e$) exhibit non-monotonic behavior.
The evolution of the spectral parameters of these five segments of S1 is shown in Figure~\ref{Fig:10} with black data points. The reduced $\chi^2$ values (0.97–1.05) confirm statistically acceptable fits in all segments. 
We also performed orbit-wise joint spectral analysis of the S2 segment, including segments for which simultaneous LAXPC and SXT data were available. We found 8 such orbits and showed their spectral-parameter evolution in Figure~\ref{Fig:10}, with red data points. We calculated the unabsorbed thermal, non-thermal, and total fluxes in the 4-25 keV energy range for individual orbits and compared them with the results from S1 segments.
We observed significant changes in the spectral parameters when the QPO turned off, i.e., when its strength decreased drastically, and the variability feature was no longer identifiable as a QPO.  

After performing the joint spectral analysis of the orbits of the S1 segment, we also examined the evolution of the QPO frequency with spectral parameters. We did not find any strong correlation between the parameters, given the uncertainties in the individual orbits. Only the power-law index shows a decreasing trend with QPO frequency, though this was not significant due to large parameter errors resulting from the relatively low SNR in the individual-orbit data. 

\begin{figure*}
\centering
\includegraphics[scale=0.5]{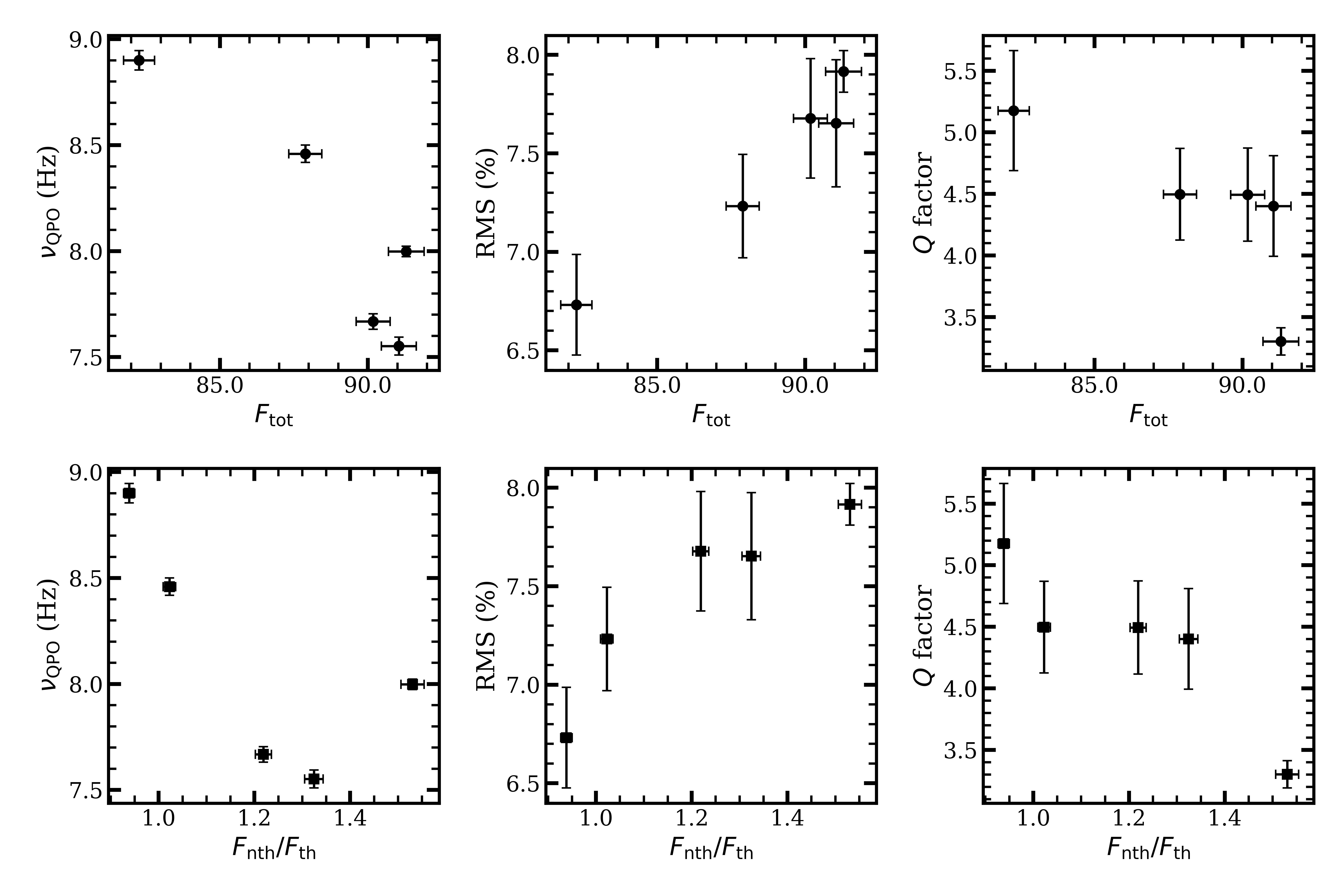}
\caption{ Evolution of the QPO parameters - centroid frequency (left), RMS (middle), and Q factor (right) - for individual orbits along with total flux (upper panels) and ratio between non-thermal to thermal flux (lower panels).  \label{Fig:11}}
\end{figure*}

We further investigated the evolution of the QPO parameters with the total flux and the flux ratio (non-thermal flux / thermal flux) in Figure~\ref{Fig:11}. We observed negative correlations between QPO frequency and both the total flux and the flux ratio, with Pearson correlation (PC) coefficients of $-0.90$ and $-0.72$, respectively. The QPO RMS showed positive correlations with total flux and the flux ratio, with PC values of 0.97 and 0.93, respectively. The $Q$ factor remained almost constant in the first four orbits, considering the error bars, and only dropped significantly at the last orbit, as seen in the second panel of Figure~\ref{Fig:10}. We have also seen negative correlations between the $Q$ factor and both the total flux and the flux ratio with PC values of -0.77 and -0.91, respectively. 

\begin{figure*}
\centering
\includegraphics[scale=0.4]{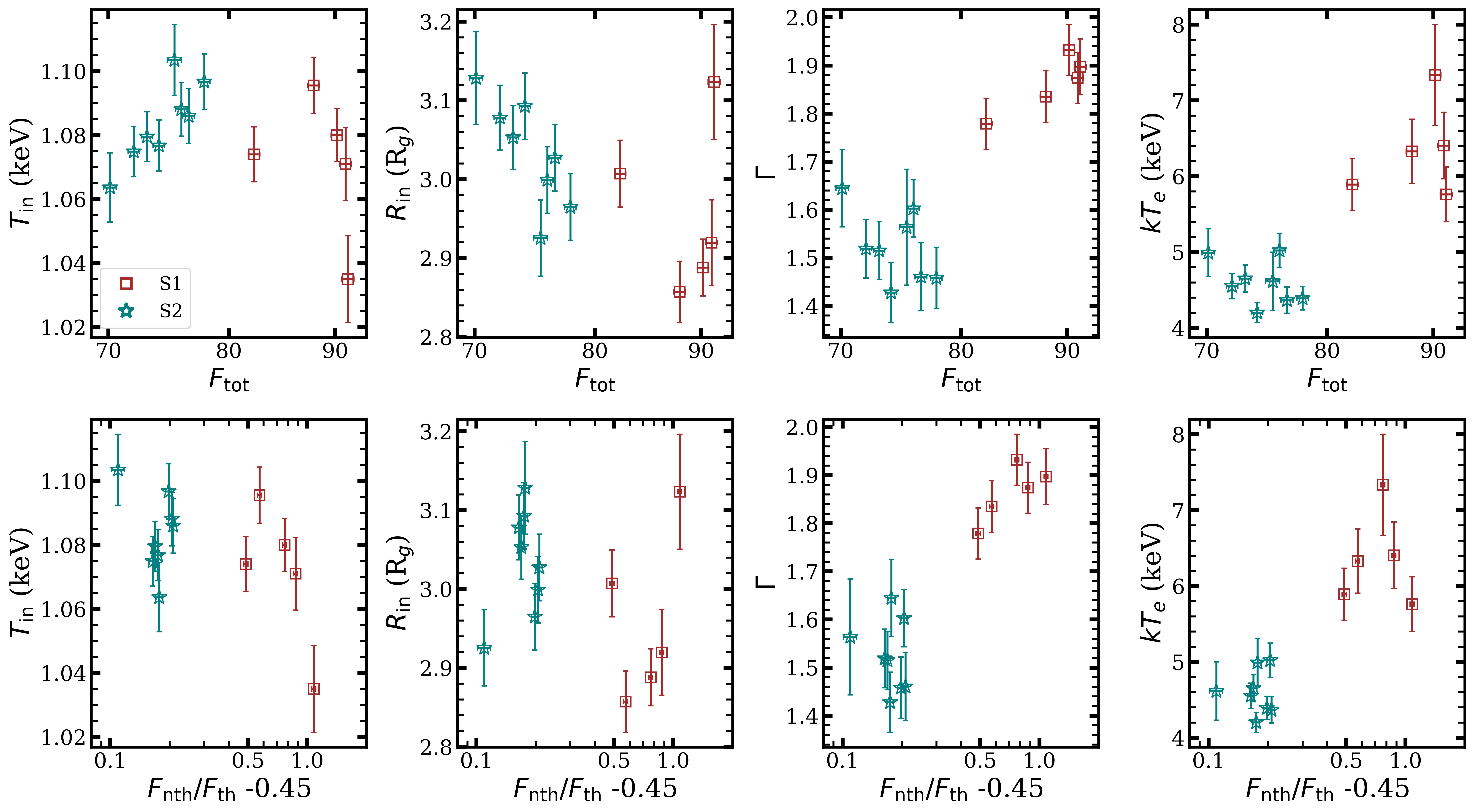}
\caption{ Evolution of the spectral parameters for individual orbits with total flux (upper panels) and ratio between non-thermal to thermal flux ratio (lower panels). The data points with the brown box marker represent the orbits of the S1 segments. The teal colored star markers represent the orbits of the S2 segment. In the x-axis of the lower panel, the ratio is plotted with the origin shifted for better visualization purposes.  \label{Fig:12}}
\end{figure*}

We also checked the evolution of the spectral parameters with total flux and flux ratio for all individual orbits considered in Figure~\ref{Fig:12}. We found that the thermal parameters do not change significantly from S1 to S2, whereas the non-thermal parameters show a substantial change, as shown in the third and fourth rows of Figure~\ref{Fig:12}. The power law index was around  1.8-2.0 for the S1 segment and decreased to nearly 1.4-1.7 for the S2 segment, while the total flux and non-thermal flux also decreased drastically. Similarly, the electron temperature also shows a small cluster around 5 keV for the S2 segment and exhibits a more dramatic evolution at slightly higher $kT_e$ values when the QPO was present in the S1 segment.
 

\section{Discussion}
\subsection{ Nature and energy dependence of the QPO}
In this work, we detected a QPO from Swift J151857.0–572147 at the start of the observation, along with a correlated low-frequency bump (LFB) near 1 Hz (Figure~\ref{Fig:2}). 
\citet{2025ApJ...987...44C} presented the monitoring of this source with \emph{HXMT} from 3rd--17th March (ending just 1 day before the \textit{AstroSat} observation) and reported the corresponding evolution of the QPO properties with the frequency varying from $\sim$3~Hz to $\sim$9~Hz.
The QPO frequency declined to $\sim$5~Hz (Q$\sim$8.4, RMS$\sim$10.4 \%) by the end of the \emph{HXMT} observation (17th March), before rising again to $\sim$8~Hz in the \emph{AstroSat} observation with an RMS of 7.8\%
and a Q factor of 3.3 
on 18th March (MJD~60387). 
The closest match to our observed QPO frequency is the $\sim$7.5~Hz QPO detected on 8th March with the ME instrument of \emph{HXMT}, which covers a similar energy range to \emph{LAXPC}, which corresponded to an RMS of 12.69\% and a Q~factor of 3.3. Though the Q factor of this QPO is similar to what was observed by LAXPC, the higher RMS could be attributed to instrumental effects or due to the epoch of the \emph{HXMT} observation, which is in a harder state compared to the \emph{AstroSat} observation. 
In this work, the strong positive correlation between the QPO and the LFB centroid frequencies (Pearson correlation coefficient of $0.98$) across individual orbital segments is entirely consistent with the idea that these components of the variability track a common underlying governing process \citep{2002nmgm.meet.2249B,1999ApJ...520..262P}. 

The energy dependence of the QPO provides important constraints on its physical origin. 
We found that both the QPO centroid frequency and the $Q$ factor increase with photon energy, and the RMS amplitude also rises monotonically from $\sim$5\% at 4--5 keV to $\sim$15\% near 10 keV, after which it saturates. This behavior in RMS, displaying a steep rise at soft X-ray energies and a plateau at higher energies, has been observed for type-C QPO frequencies above $\sim 2$ Hz in several black hole transients, including 
XTE J1859+226, GRS 1915+105,  XTE J1550–564, Swift J1727.8–1613 and H 1743–322 \citep{2004A&A...426..587C, refId0, 2010ApJ...714.1065R,2024ApJ...968..106Z,2013MNRAS.433..412L}. 
For Swift J151857.0–572147, \citet{2025ApJ...987...44C} reported that the QPO RMS increases from the LE (low energy) to the ME (mid energy) range of \emph{HXMT} (consistent with our results), before decreasing in the HE (high energy) range. 
The increasing RMS with energy, along with consistently high values above 10 keV, likely indicates that the Comptonized component is the primary driver of the QPO variability. This is consistent with Lense--Thirring precession models \citep{1998ApJ...492L..59S, 10.1111}, where the modulation of the hot inner flow produces stronger variability at energies dominated by inverse Comptonization. 
 Along with the QPO, the LFB also shows an increasing energy dependence, indicating that the physical processes modulating the QPO emission also affect the LFB component.

The hard time lags of the QPO,  rising monotonically with energy, 
further reinforce the Comptonization picture. Such positive lags, where the hard photons lag the soft photons at the QPO frequency, are a characteristic signature of X-ray reverberation within a Comptonizing corona
\citep{2014A&ARv..22...72U,2019Natur.565..198K}. The relatively modest magnitude of the lags ($<$12 ms) is consistent with a compact corona \citep{2021MNRAS.503.5522K,2019Natur.565..198K,2022MNRAS.515.2099B,2026MNRAS.tmp..912R}, with moderate electron temperature ($kT_e \sim 6$--8 keV) inferred from the spectral fitting. Within the Lense--Thirring precession framework, the sign of the QPO phase lag carries inclination information: low-inclination sources are expected to exhibit hard (positive) lags at the type-C QPO frequency ($>$ 2 Hz), while high-inclination sources tend to show soft (negative) lags \citep{2013ApJ...778..165V, 2016MNRAS.458.3655V,2013ApJ...778..136P,2026MNRAS.tmp..912R}. The persistently hard lags observed here are therefore consistent with Swift J151857.0--572147 being viewed at a relatively low inclination. This is in agreement with the low disk inclination independently constrained via continuum spectral fitting by \citet{2024ApJ...975..257M}. 
This consistency between the timing-based and continuum-spectroscopy-based inclination estimates demonstrates that QPO time-lag analysis can serve as an independent, geometry-sensitive probe of the viewing angle in black hole low-mass X-ray binaries, complementing traditional spectral methods.

\subsection{ Disappearance of the QPO and the low-frequency bump}

The most notable result of this study is the abrupt disappearance of the type-C QPO together with the correlated LFB after the first six orbital segments (S1), with no recovery detected in the subsequent nine orbits (S2). The dynamic power spectra clearly show a QPO near 8.0 Hz and an LFB near 1 Hz exclusively during S1, and both features vanish simultaneously at the S1--S2 boundary. The simultaneous disappearance of these two correlated timing features is significant: it implies that whatever physical mechanism sustains the QPO also impacts the correlated low-frequency variability and that both are quenched by the same transition in the accretion geometry. 

The simultaneous disappearance of the type-C QPO and the LFB at the S1--S2 boundary finds direct analogs in several other BH-LMXBs, supporting the interpretation that both features share a common physical origin in the hot inner flow. \citet{2020ApJ...891L..29H} reported that in MAXI~J1820+070, an abrupt type-C to type-B QPO transition was accompanied by a simultaneous drop in broadband X-ray variability and a brief 7--12 keV flare, followed $\sim$2--2.5 hr later by a strong radio flare marking the launch of superluminal ejecta.
A related behavior was reported in GX~339$-$4 by \citet{Buisson2025FlipflopQC}, who found that states with a strong QPO are accompanied by enhanced very low-frequency noise (0.01--0.1 Hz), whereas the dim state, in which the QPO is absent, shows significantly weaker variability at these frequencies, indicating that the low-frequency variability and the QPO are co-produced by the same accretion structure. Most recently, \citet{2025A&A...699A...9J} reported the simultaneous suppression of the low-frequency broadband noise component coincident with the type-C QPO disappearance in Swift J1727.8$-$1613, directly mirroring what we observe in Swift J151857.0$-$572147. 
Taken together, these results consistently indicate that the type-C QPO and the associated low-frequency broadband variability are co-generated in the Comptonizing inner flow, and that their simultaneous quenching signals a rapid restructuring of the coronal geometry at the harder-to-softer state boundary.

In this work, the hardness-intensity diagram for \src shows that the QPO disappearance coincides with a distinct spectral-state transition. During S1, the source traced a zig-zag pattern in the HID with hardness $\sim$0.20--0.24, consistent with the behavior in the intermediate state  \citep{2010LNP...794...53B}. By orbit O6, the source begins to move to lower intensity and softer spectral states, and in S2 it occupies a clearly separated, softer region of the HID with hardness $\sim$0.17--0.20 and reduced intensity ($\sim$680--760 counts s$^{-1}$). 
Using the hue values calculated following \citet{2015MNRAS.448.3339H}, we observed the hue changes dramatically from 177$^\circ$ in S1 to 13$^\circ$ in S2, marking a transition from a harder towards a softer spectral state \citep{2015MNRAS.448.3339H,2026MNRAS.tmp..912R}. 
 During our observation, the inner disk temperature is measured as 1.08 ~keV, while the power-law index is 1.86, indicating that the source was possibly in the SIMS state, or it could be transitioning from the SIMS towards the soft state. Earlier reported works also suggest this source to be in the soft or soft-intermediate state before the start of the \emph{AstroSat} observation \citep{2024ATel16519....1D,2024ApJ...975..257M,2025ApJ...987...44C}. 
Taken together, these results indicate that the S1--S2 transition observed here reflects a harder-to-softer state transition, possibly near the tail end of the SIMS state.
This transition is exactly the type of evolution during which QPO disappearance has been observed in other black hole transients during the outburst \citep{2006ARA&A..44...49R, 2012A&A...542A..56N,2004A&A...426..587C}.

The disappearance of the QPO has been observed in other sources as well, but the reason is still not clearly understood. 
Our spectral analysis results provide a physically coherent explanation for this behavior. Across the S1--S2 transition, the disk temperature remains the same 
, along with a slight increase in the thermal normalization  considering the 1$\sigma$ error bar (Table \ref{Tab:1}), representing a modest outward movement of the apparent inner disk radius and a reduction in spectral hardening. 
In contrast, the photon index changed from 1.86 to 1.45, the electron temperature dropped from 6.42 keV to 4.46 keV, and the Comptonization normalization fell substantially (Table ~\ref{Tab:1}). 
Correspondingly, the non-thermal flux decreases from $\sim$48 to $\sim$28 (in units of $10^{-10}$\,erg\,s$^{-1}$\,cm$^{-2}$), indicating a significant reduction in Comptonized emission. The simultaneous decrease in coronal temperature, non-thermal flux, and hardness ratio together indicates an overall weakening and cooling of the corona, in spite of the slight hardening of the non-thermal photon index. 
This is further supported by our orbit-wise analysis, where the thermal spectral parameters show a similar scattered distribution across the orbits of both S1 and S2 (Figure~\ref{Fig:11}), but the non-thermal parameters form two distinct clusters corresponding to the two segments. 

This behavior mirrors what has been reported across several BH-LMXBs: in GX 339-4, \citet{2023MNRAS.521.3570Y} found that the inner accretion disk radius remained essentially constant during the QPO transitions. 
For H 1743-322, \citet{2021ApJ...911..127S} demonstrated that the inner disk radius was stable at $\sim$2--9\,$r_{\rm g}$ throughout the transition, while the power-law photon index varied significantly, implying that the observed evolution is governed by the corona or jet rather than the variation of the disk truncation radius.
Such stability of the thermal disk parameters across a QPO transition has been previously reported \citep{2012A&A...541A...6S,2009MNRAS.392..992D}.
In MAXI J1348--630, \citet{2021MNRAS.505.3823Z} found that the disk flux fraction actually decreases while the Comptonized flux increases when the QPO appears, demonstrating that the disk is not the primary driver of QPO transitions. A qualitatively similar hardening of the non-thermal component coincident with QPO suppression has been reported in XTE J1550--564 \citep{2003ApJ...595.1032R} and in the 2021 outburst of GX 339--4 \citep{2023MNRAS.526.4718M}, where the photon index was observed to decrease as the source transitioned across the hard-to-soft state boundary and the QPO weakened. In Swift J1658.2--4242, \citet{2019ApJ...879...93X} similarly documented that both disk and power-law fluxes decrease during the QPO transition.
These observed evolution of the spectral parameters collectively indicate that the corona rapidly settles into a cooler, stable configuration appropriate to the softer state \citep{2023MNRAS.525..854M}, establishing that it is the corona that primarily governs both QPO generation and its eventual quenching, consistent with the broader picture established across multiple BHXBs \citep{2009MNRAS.392..992D, 2003ApJ...595.1032R, 2021MNRAS.505.3823Z, 2020A&A...641A.101B}. Given that the QPO RMS is dominant at high energies, where the primary contribution is from the Comptonized component \citep{2021MNRAS.505.3823Z, 2020MNRAS.492.1399K}, the collapse of the corona naturally explains the disappearance of the QPO. This combination of a collapsing corona alongside a stable disk is the defining spectral signature of the hard to soft state transition and has been consistently reported in GX 339--4 \citep{2009MNRAS.392..992D, 2021MNRAS.508..287S}, XTE J1550--564 \citep{2003ApJ...595.1032R}, and Swift J1658.2--4242 \citep{2020A&A...641A.101B}.  

The disappearance of the QPO can be explained within the Lense-Thirring framework, the type C QPO frequency scales as $f_{\rm QPO} \propto r_{\rm tr}^{-3}$, where $r_{\rm tr}$ is the truncation radius of the hot inner flow \citep{2019NewAR..8501524I}.
For the observed QPO frequency of $\sim$8\,Hz, this relation implies a truncation radius of $r_{\rm tr} \sim 10$--$15\,r_g$, consistent with the expected radial extent of the geometrically thick Comptonizing flow in the hard-intermediate state \citep{2025MNRAS.543.1748M}. 
If the QPO disappearance occurs due to the quenching of the hot inner flow as the inner disk truncation radius moves from  $\sim 15\,r_g$ down to the ISCO at $r_{\rm ISCO} \sim 2.5\,r_g$, then, the upper limit of relevant viscous timescale \citep{1981ARA&A..19..137P,2001MNRAS.321..759C} for such a process is estimated to be nearly 2600 s (assuming $M_{\rm BH} = 10\,M_\odot$, $r_{\rm tr} = 15\,r_g$, $\alpha = 0.01$, and $H/r = 0.01$).
This is in good agreement with the observed gap of $\sim$1860\,s ($\sim$0.52\,hr) between the last orbit of S1 and the first orbit of S2. However, it is to be noted that from our spectral analysis result, we do not have such an indication of a change in inner disk radius or disk temperature, except for the thermal flux evolution.

Although so far no radio observations have been reported around the time of our \emph{AstroSat} epoch, observation with the \emph{Australia Telescope Compact Array (\textit{ATCA})} shows a major radio flare peaking at $\sim 1550$ mJy (9 GHz) on 9th March \citep{2024ATel16518....1C}. Concurrently, the \emph{Australian SKA Pathfinder (\textit{ASKAP})} flux density displayed a rise to $\sim$50 mJy at 887.5 MHz on 11th March 2024 from a pre-outburst limit of $\lesssim 3$ mJy (21st February 2024) and then fell to $\sim$ 7 mJy on 27th March 2024 \citep{2024ATel16617....1A}. This places Swift J151857.0–572147 in a post-flare, radio-fading state during the \emph{AstroSat} observation, which was on 18th March 2024. This temporal sequence, a bright radio flare peaking $\sim$7 days prior, followed by rapid radio fading contemporaneous with our observation, is entirely consistent with the standard picture in which a jet ejection event precedes the change in the QPO behavior by a few days, as seen in MAXI J1820+070 \citep{2020ApJ...891L..29H, 2025MNRAS.538.1143L,2023MNRAS.525..854M}, Swift J1727.8--1613 \citep{2025A&A...699A...9J, 2025ApJ...984L..53W}, and the multi-source sample of \citet{2016MNRAS.460.4403R}. In those sources, the radio flare, associated with the jet ejection, peaks within hours to a few days of the QPO transition and turns off. A similar result was reported in GRS 1915+105 by \citet{2001A&A...370L..17V}, who found that during a $\sim$450 mJy radio flare, the 0.5--10 Hz QPO disappeared and the Comptonized component weakened drastically, leading them to conclude that the corona, responsible for both the hard X-ray emission and the QPO, was ejected during the event. Consistently, in this study, the X-ray signatures we observe - simultaneous suppression of the type-C QPO and LFB, coronal cooling, and non-thermal flux decrease - therefore indicate an evolved coronal state of the X-ray binary consistent with a near-quenched jet post the radio flare. 
Contemporaneous multi-wavelength and high-sensitivity monitoring of future outbursts will be essential to probe the underlying QPO turn-off mechanism and its connection with the corona and jet evolution. 

\section*{Acknowledgements}
This study incorporates data collected through the \emph{AstroSat} mission, which is accessible to the public via the ISRO Science Data Archive for the \emph{AstroSat} Mission. The Indian Space Science Data Centre (ISSDC), ISRO, facilitates user access to the data (https://astrobrowse.issdc.gov.in/
astro$\_$archive/archive/Home.jsp). We are grateful to the LAXPC team for providing the data and the required software tools for the analysis. 
B. R. acknowledges the financial support of the Council of Scientific and Industrial Research (CSIR), India, under the CSIR-SRF Direct fellowship, File No. 09/1022(25279)/2025-EMR-I.

\section*{DATA AVAILABILITY}
Data analyzed in this work is publicly available on the Indian Space Science Data Center (ISSDC) website (https://astrobrowse.
issdc.gov.in/astro\_archive/archive/Home.jsp).



\bibliographystyle{elsarticle-harv} 
\bibliography{ref}







\end{document}